\documentclass[%
 reprint,
 amsmath,amssymb,
 aps,
]{revtex4-2}

\usepackage{graphicx}
\graphicspath{{./figs/}}
\usepackage{caption}
\usepackage{subcaption}
\usepackage{dcolumn}
\usepackage{bm}

\begin{document}

\preprint{APS/123-QED}

\title{A two-qubit entangling gate based on a two-spin gadget}

\author{Rui Yang}

\affiliation{%
 Institute for Quantum Computing, University of Waterloo\\
}%



%



\date{\today}

\begin{abstract}
The faster speed and operational convenience of two-qubit gate with flux bias control makes it an important candidate for future large-scale quantum computers based on high coherence flux qubits. Based on a properly designed two-spin gadget which has small gaps during the evolution of energy levels, we build a CNOT-equivalent gate which can reach a fidelity larger than $99.9\%$ within 40ns. Moreover, we also use the Schrieffer-Wolff Transformation to translate the spin model Ising coefficients schedule to circuit model flux bias schedule for realistic flux qubit circuits coupled by a tunable rf-squid. The two-qubit entangling gate scheme introduced here is suitable for realizing efficient two-qubit gates in the large scale flux qubit systems dominated by inductive couplings. Comparing with the current gate-based quantum computation systems dominated by capacitive couplings, it can resolve the conflict between the speed and a high coherence.

\end{abstract}

\maketitle


\section{\label{sec:level1}Introduction:\protect
}

The field of quantum computation is developing fast these days. Nowadays, researchers can routinely control tens of qubits with a high precision. The quantum information community has entered the so-called NISQ era and strong evidence for quantum advantage has been demonstrated in systems based on transmon qubits\cite{Arute_2019, USTC2021}. A two-qubit entangling gate along with a single qubit gate can be used to realize arbitrary unitary operations needed for quantum computation. A high fidelity and high efficiency two-qubit entangling gate is a key element in quantum computation, which will reduce overhead for performing quantum error corrections for large scale quantum computation\cite{huang2021ultrahigh}. The most efficient two-qubit gates belong to the CNOT gate and iSWAP gate families\cite{huang2021ultrahigh}. With the help of arbitrary single-qubit gates, it will only need up to 3 uses of CNOT or iSWAP gate to realize arbitrary two-qubit unitary evolution\cite{huang2021ultrahigh}. The CNOT and iSWAP gates therefore draw much attention from the gate-based quantum computation community. A key limiting factor for realizing a high fidelity gate is the coherent time of qubits. So far, transmon has been dominating the field of superconducting quantum computation, which has a simple structure and a decent coherence time (10s of microseconds)\cite{Arute_2019}. However, to reach a fidelity required by the quantum error correction, qubits with even higher coherence are favored. Recent studies shows that, comparing with transmon, the fluxonium qubit has enhanced coherence. It is therefore a promising better replacement for transmon qubit to realize an error-correctable quantum computer. The high coherence of fluxonium qubit is rooted in the suppressed transition matrix element of the charge operator of the fluxonium\cite{bao2021fluxonium}. However, for gates based on capacitive coupling (which is the dominating design for gate-based quantum computation), this suppression will inevitably slows down the gate speed. Inductive coupling, however, doesn't suffer from the above limitation\cite{bao2021fluxonium}. Moreover, for large scale quantum computation, a tunable coupler is crucial for maintaining high-fidelity for parallel operations\cite{bao2021fluxonium}.

In terms of operations, there are two specific ways for making two-qubit gates -- one way is to use only flux-bias control, the other way is to use parametric driving. In the first way, anticrossings in the levels (tunable by flux bias) can be used for creating desired gate\cite{Arute_2019}. For example, a CZ gate has been realized in transmon qubit systems by the anticrossing of $|11\rangle$ and $|02\rangle$ levels\cite{CZ1, DiCarlo_2009}. For two frequency-tunable transmon qubits coupled capacitively. When biasing the flux threading the SQUID loop in the transmon qubit, the $|11\rangle$ and $|02 \rangle$ levels will come close to each other and forming an anticrossing due to the coupling between the two qubits. Starting with the two-transmon system biased far from the anticrossing point, then, biasing the flux bias line to make a short itinerary to the anticrossing point and coming back to the bias point, a phase will be picked up\cite{CZ2, CZ3}. With this mechanism, we can make a CZ gate, with the evolved final state for one of the two transmons depending on the initial state of the other transmon qubit\cite{CZ2, CZ3}. The 2nd method of making two-qubit gate is via parametric driving on one of the qubit while parking qubits at fixed flux bias\cite{Hong,Vinay}. For the parametric scheme, one of the two qubits will be parked at a fixed flux bias (e.g. flux sweet spots), and driving on the other qubit will introduce cross-resonance on the qubit parked at a fixed bias. Comparing these two methods, the advantage of using only flux biasing to make a two-qubit gate is the stronger interaction and faster speed, since it can use the whole interaction between qubits\cite{Hong,Vinay}. The disadvantage of this scheme is that it is more susceptible to dephasing since the excursion could make the system move away from sweet spots\cite{Hong,Vinay} (first order flux insensitive points) during gate operation. This problem can be mitigated for new generation of flux qubit with much higher coherence and optimized waveform to shorten the excursion away from sweet spots. For the parametric gate, the interaction is much weaker and the speed is much more slower, since the interaction is from sideband resonance\cite{Hong,Vinay}. In general, both methods have pros and cons and both can lead to useful gates. The faster speed and operational convenience of two-qubit gate with flux bias control makes it an important candidate for future large-scale quantum computers based on high coherence flux qubits such as fluxonium.

To realize two-qubit gates in large-scale flux qubit systems without the speed limitation caused by the capacitive coupling, we conceived an efficient two-qubit gate (CNOT-like gate) scheme using only flux-bias control, for two flux qubits inductively-coupled with a tunable coupler. Specifically, by using a properly designed time-dependent annealing Hamiltonian from our previous study, we constructed a two-spin gadget which has small gaps in the evolution of energy levels. Using one small gap (anticrossing) in our gadget, we can build a two-qubit entangling gate which rotates the 1st spin only when the 2nd spin is in the ground state. Similar to the CZ gate used in transmons, the two-qubit gate here also uses only flux bias control. The two-qubit gate is realized by flux biasing the dimer system close to an anticrossing point and move back to the initial bias point. Moreover, we also studied how to realize such a spin model with realistic capacitively-shunted flux qubit(CSFQ) circuits coupled by a tunable rf-squid. We use the Schrieffer-Wolff Transformation\cite{translator1, translator2, translator3} to translate the spin model Ising coefficients schedule to circuit model flux bias schedule\cite{translator1, translator2}. The low-level dynamics of the circuit model successfully reproduces the desired spin model evolution. Therefore, our entangling gate are realizable with realistic Josephson circuits.

\section{Spin model}

\subsection{A two-spin gadget(time-dependent Hamiltonian and level evolution)}

Our two-qubit gate is based on the time-dependent Hamiltonian of a two-spin gadget, which reads\citep{translator1}:
\begin{align}
	H(s) &= \gamma_{d1}(s)h^x_1\sigma^x_{1} + \gamma_{d2}(s)h^x_2\sigma^x_{2} \nonumber \\
	&+ \gamma_p(s)[h^z_1 \sigma^z_{1} + h^z_2 \sigma^z_{2} + J \sigma^z_{1}\sigma^z_{2}]
\end{align}

where $h_1^z$, $h_2^z$, $h_1^x$, $h_2^x$ and $J$ are constant and satisfying  $0<h_1^z< h_2^z$, $h_1^z<J< h_2^z$, $h_1^x < h_2^x$, and $\gamma_p(s)$, $\gamma_{d1}(s)$, and $\gamma_{d2}(s)$ are schedule functions in the range $[0, 1]$, with $s=t⁄t_f$ where $t_f$ is the total annealing/evolution time. $\gamma_d1 (s)$, $\gamma_d2 (s)$, $\gamma_p (s)$ are dimensionless factors.

To make two small gaps, the annealing schedule of this problem is divided into two parts, in order to get a Hamiltonian giving an annealing evolution for a dimer – the horizontal part of the Hamiltonian ( $\lambda_d1 (s) h_1^x \sigma_1^x+ \lambda_d2 (s) h_2^x \sigma_2^x$  ) gradually vanishes and the longitudinal part of the Hamiltonian ( $ \lambda_p (s)(h_1^z \sigma_1^z+ h_2^z \sigma_2^z+J\sigma_1^z \sigma_2^z )$ ) gradually emerges from 0.

In the first part of the annealing evolution, for $s\in [0, s_1]$, the annealing schedule reads:

\begin{equation}
	0 \leq s \leq s_1 :
	\begin{cases} 
		\gamma_{d1}(s) = \left( \frac{\Delta_\text{min}^{(1)}}{2h^x_1} - 1 \right)\frac{s}{s_1} + 1 \\
		\gamma_{d2}(s) = 1 \\
    	\gamma_p(s) = 0  
   \end{cases}
\end{equation}
where $\Delta_\text{min}^{(1)}$ is the first small gap in this problem occurring at $s=s_1$, since for this initial part of the anneal the gap is always $2\gamma_{d1}(s)h^x_1$.

In the second part of the annealing evolution, for $s\in (s_1, 1]$, the annealing schedule reads:

\begin{equation}
	s_1 < s \leq 1 :
	\begin{cases} 
		\gamma_{d1}(s) = \frac{\Delta_\text{min}^{(1)}}{2h^x_1} \frac{s - 1}{s_1 - 1} \\
		\gamma_{d2}(s) = \frac{s - 1}{s_1 - 1} \\
    	\gamma_p(s) = \frac{s - s_1}{1 - s_1}  
   \end{cases}
\end{equation}

Since $h_1^{x} << h_2^{x}$, we can approximate the gap of the system as

\begin{equation}
	\Delta^{(2)}(s) \approx \sqrt{[\tilde{\Delta}(s)]^2 + [2\gamma_{d1}(s)h^x_1]^2},
\end{equation}
where 
\begin{align}
	\tilde{\Delta}(s) &= \sqrt{\gamma_p^2(s)(h^z_2 + J)^2 + [\gamma_{d2}(s)h^x_2]^2} \nonumber \\
	&- \sqrt{\gamma_p^2(s)(h^z_2 - J)^2 + [\gamma_{d2}(s)h^x_2]^2} \nonumber \\
	&- 2\gamma_p(s)h^z_1
\end{align}
is the gap of the system in the absence of $h^x_1$.
If $h^z_1 < J, h^z_2$ (as we assumed earlier) then there exists $s^* \in (s_1, 1]$ for which $\tilde{\Delta}(s^*)=0$, and therefore the system reaches its second small gap of $\Delta^{(2)}_\text{min} = \Delta^{(2)}(s^*) \approx 2\gamma_{d1}(s^*)h^x_1$ for this part of the anneal.

\begin{figure}
     \centering
     \begin{subfigure}[t]{1\columnwidth}
         \caption{}
         \centering
         
         \includegraphics[width=1\columnwidth]{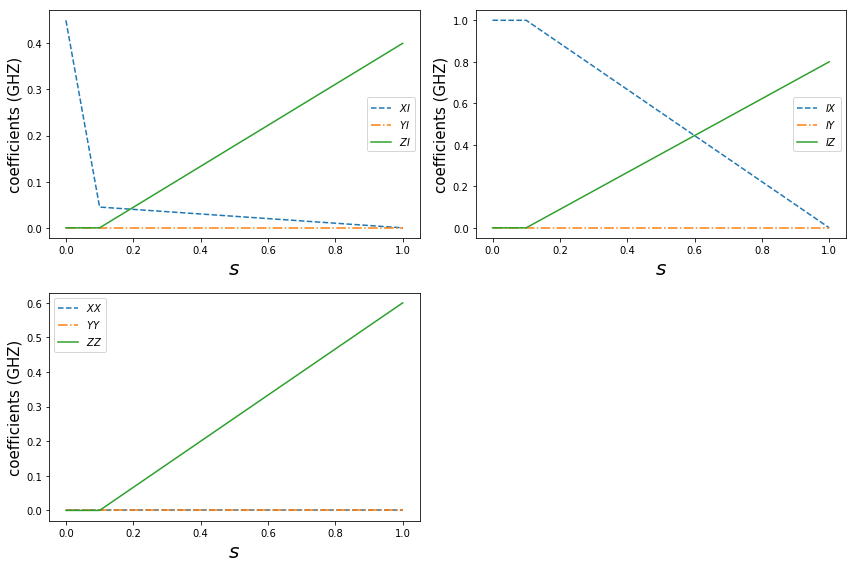}%
         \label{fig:Ising_schedules}
     \end{subfigure}
     \hfill
     \begin{subfigure}[t]{1\columnwidth}
         \caption{}
         \centering         
         \includegraphics[width=1\columnwidth]{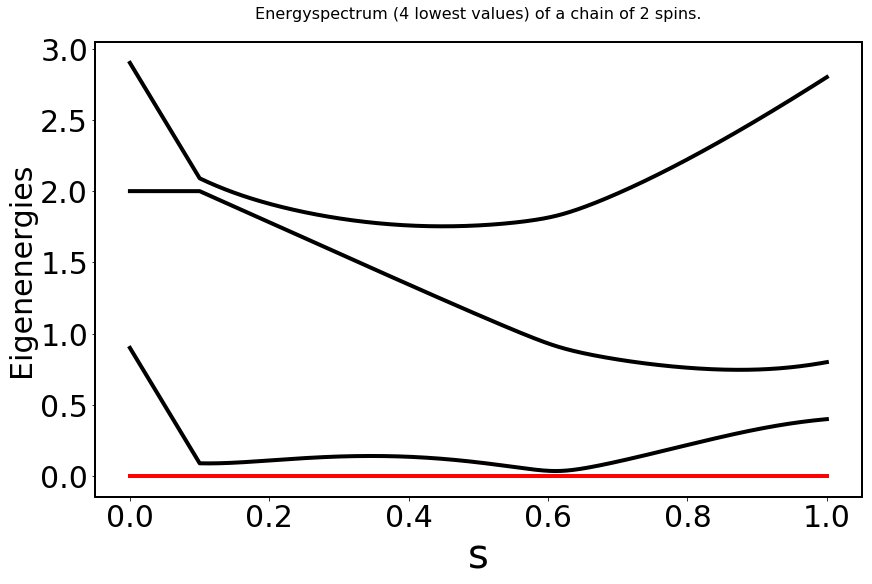}%
         \label{fig:level_evolution_dimer}
     \end{subfigure}

        \caption{Dimer (a) Ising coefficients as a function of reduced time s (Ising schedule) for Hamiltonian of the spin model of our two-spin gadget (see equations (1) to (3)) (b)The evolution of 4 energy levels of our two-spin gadget caused by the time-dependent Hamiltonian of our two-spin gadget }
        \label{fig:dimer_figure}
\end{figure}

The evolution of energy levels can be found in Fig.\ref{fig:dimer_figure}, (a) schedule, (b) levels evolution.

\subsection{Coherent oscillation}

\begin{figure}[t]
	\begin{center}
		\includegraphics[width=1\columnwidth]{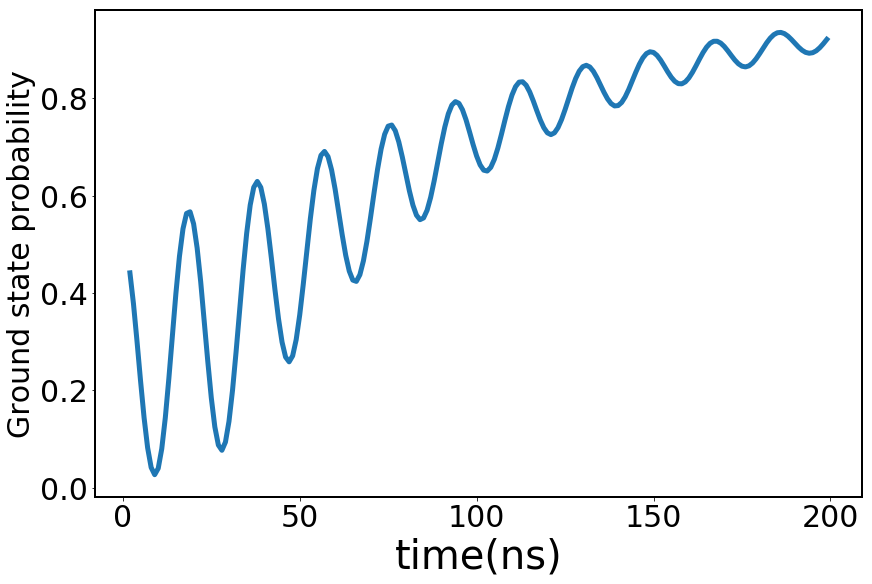}%
    \end{center}
    \caption{The final ground state population as a function of total annealing time for our two-spin gadget, which shows a coherent oscillation behavior.The oscillation origins from the interference between different paths connecting the initial state to the ground state of the final Hamiltonian}\label{fig:oscillation_dimer}
\end{figure}

Now, let's inspect the time evolution for the gadget. Our simulation reveals that the final ground state probability is a function of total annealing time $t_f$, which shows an oscillation (see Fig.\ref{fig:oscillation_dimer}). Looking at the time evolution of energy levels will shed light on the nature of this oscillation. In Figure~\ref{fig:dimer_figure}, we can see the evolution of the 4 energy levels, we can see two anticrossings occur in the evolution. To reach the final ground state, there are different paths in the evolution: the system could jump onto the 1st excited state at the 1st anticrossing and jumps back to the ground state at the 2nd anticrossing, or, the system could just stay at the ground state. The oscillation is caused by the interference between different paths reaching the ground state of the final hamiltonian. The two anticrossings serve as two beam splitters and the interference between the paths depends on $t_f$. This phenomenon has been reported and studied in a single qubit scenario\cite{Munoz-Bauza2019}. As far as we know, our dimer DQA result\cite{translator1} is the first case of realizing this oscillation behavior in a dimer system.

Previous study shows that this type of diabatic coherent oscillation will be damped once the coupling with environment kicks in. A study on a single qubit scenario demonstrates that decoherence caused by coupling with a thermal bath will lead to the damping of the oscillation\cite{Munoz-Bauza2019}. The oscillation observed in our dimer gadget was also revealed to be influenced by the decoherence caused by the environment, as indicated in our previous publication\cite{translator1}.

\subsection{Two-qubit gate}

\begin{figure}
     \centering
     \begin{subfigure}[b]{0.8\columnwidth}
         \centering
         \caption{}
         \includegraphics[width=1\columnwidth]{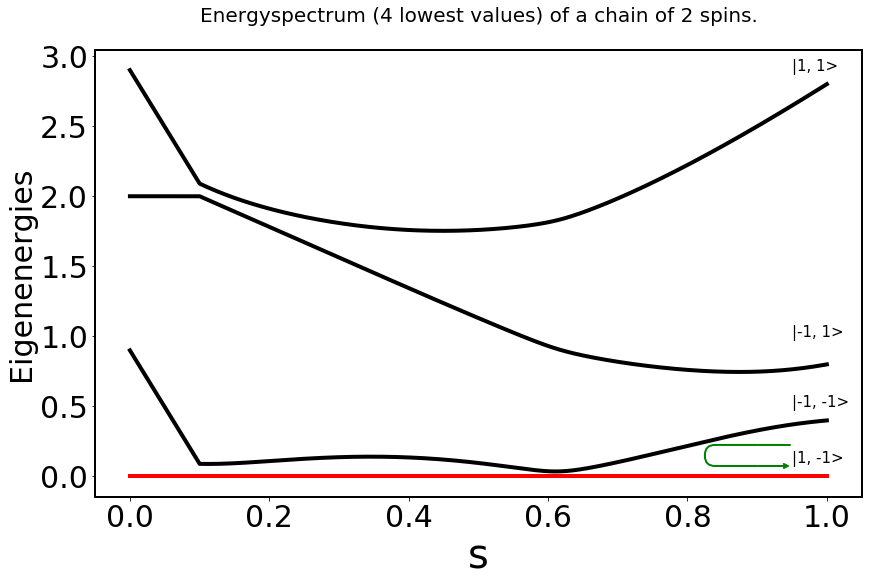}%
         \label{fig:gate_waveform_levels}
     \end{subfigure}
     \hfill
     \begin{subfigure}[b]{0.8\columnwidth}
         \centering
         \caption{}
         \includegraphics[width=1\columnwidth]{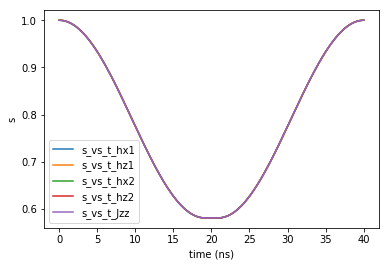}
         \label{fig:gate_waveform}
     \end{subfigure}
     \hfill
     \begin{subfigure}[b]{0.8\columnwidth}
         \centering
         \caption{}
         \includegraphics[width=1\columnwidth]{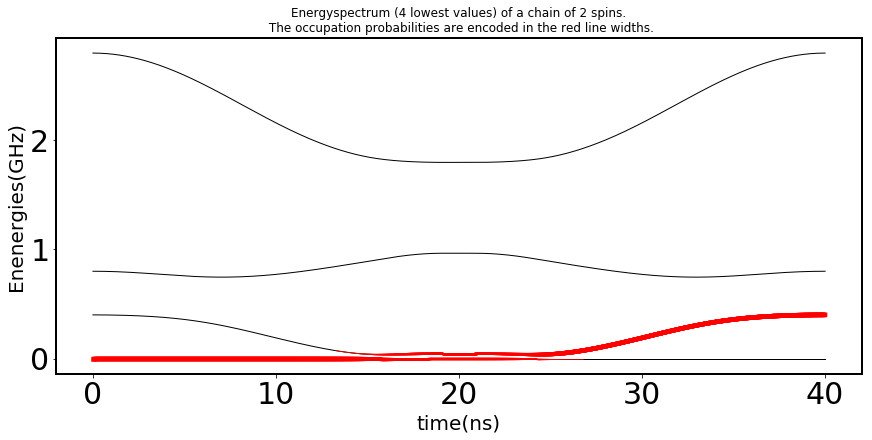}
         \label{fig:gate_waveform}
     \end{subfigure}

        \caption{Two-qubit gate based on the anticrossing of the levels evolution. (a)The green arrow shows the concept of how the two-qubit gate is created by moving close to the anticrossing point in the level evolution. (b)The Ising coefficients as a function of time for the two-qubit gate can be indicated by the reduced time s of the DQA gadget as a function of time, where each s corresponds to a set of Ising coefficients (c)The evolution of energy levels evolution and populations for the two-qubit gate }
        \label{fig:two-qubit gate}
\end{figure}
Now, let's inspect the previous gadget in a different perspective. Starting with s=1 and biasing the dimer system close to the 2nd anticrossing point and move back to the initial bias point at s=1, we can make a two-qubit gate based on the gadget (see Fig.\ref{fig:two-qubit gate}). Fig.\ref{fig:two-qubit gate} shows the spin model results. In Fig.\ref{fig:two-qubit gate}(a), the green arrow shows the variation of s. The s starts at 1 at beginning, then, s decreases until reaches around 0.6, then, s increases again until reaches 1 again. Fig.\ref{fig:two-qubit gate}(b) shows the s as a function of time in this two-qubit gate operation. Fig.\ref{fig:two-qubit gate}(c) shows the population swap between the ground and 1st excited levels of the two-qubit system when the control qubit is in its ground state. Therefore, the two-qubit gate rotates the 1st spin only when the 2nd spin is in the ground state. After further studying the action of the unitary, we found it has this general form:
\begin{align}
U_{gate} = \begin{bmatrix}
\cos(\eta)\exp(i\nu_1) & \sin(\eta)\exp(i\theta) & 0 & 0\\
\sin(\eta)\exp(i\theta) & \cos(\eta)\exp(i\nu_2) & 0 & 0\\
0 & 0 & \exp(i\phi_1) & 0\\
0 & 0 & 0 & \exp(i\phi_2)
\end{bmatrix}
\end{align}

The elements of this unitary matrix depend on the system parameters (of the two-spin gadget) and the waveform parameters. After adjusting the waveform parameters of the Ising coefficients, we can make the unitary matrix approximate a more regular form with a high accuracy:

\begin{align}
U_{gate} = \begin{bmatrix}
0 & i & 0 & 0\\
i & 0 & 0 & 0\\
0 & 0 & -1 & 0\\
0 & 0 & 0 & -1
\end{bmatrix}
\end{align}

Therefore, adding a single-qubit $\pi/2$ pulse can make our two-qubit gate a CNOT gate. This two-qubit gate is similar to CZ gate used in transmons, the two-qubit gate here is also realized by biasing the dimer system close to the anticrossing point and move back to the initial bias point to make different computational basis to pick up different phases or unitary transformations during the process. 

\begin{figure}[t]
	\begin{center}
		\includegraphics[width=1\columnwidth]{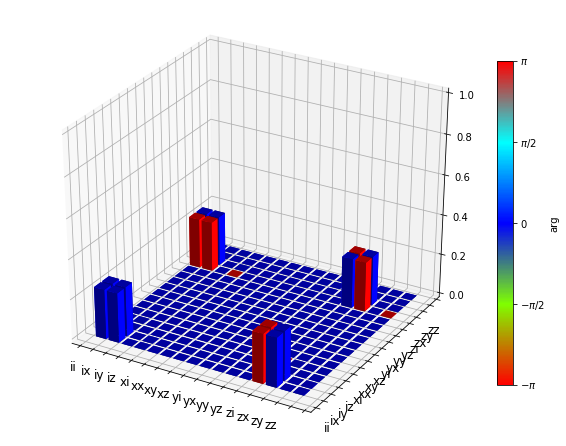}%
    \end{center}
    \caption{Quantum process tomography (QPT) of the two-qubit gate plus a single-qubit $\pi/2$ gate. The corresponding fidelity of our two-qubit gate is $>99.9\%$.}\label{fig:gate_barplot}
\end{figure}

The corresponding optimized waveform of flux bias can be seen in Fig.\ref{fig:two-qubit gate}(b). Within 40ns, the fidelity of this CNOT-like gate can reach more than $99.9\%$ with a simple flattop cos waveform. The corresponding s vs time function for spin model is shown in Fig.\ref{fig:two-qubit gate}(b). The quantum process tomography for a CNOT gate realized with our two-qubit gate plus a $\pi/2$ gate can be found in Fig.\ref{fig:gate_barplot}. We can see the QPT result is very close to an ideal CNOT gate.

\section{Circuit Model}

We also studied how to realize such a spin model with realistic capacitively-shunted flux qubit(CSFQ) circuits coupled by a tunable rf-squid coupler\cite{translator1}. We used the Schrieffer-Wolff Transformation to translate the spin model Ising coefficients schedules to circuit model flux bias schedules\cite{translator1}.

\begin{figure}
     \centering
     \begin{subfigure}[b]{0.8\columnwidth}
         \centering
         \caption{}
         \includegraphics[width=1\columnwidth]{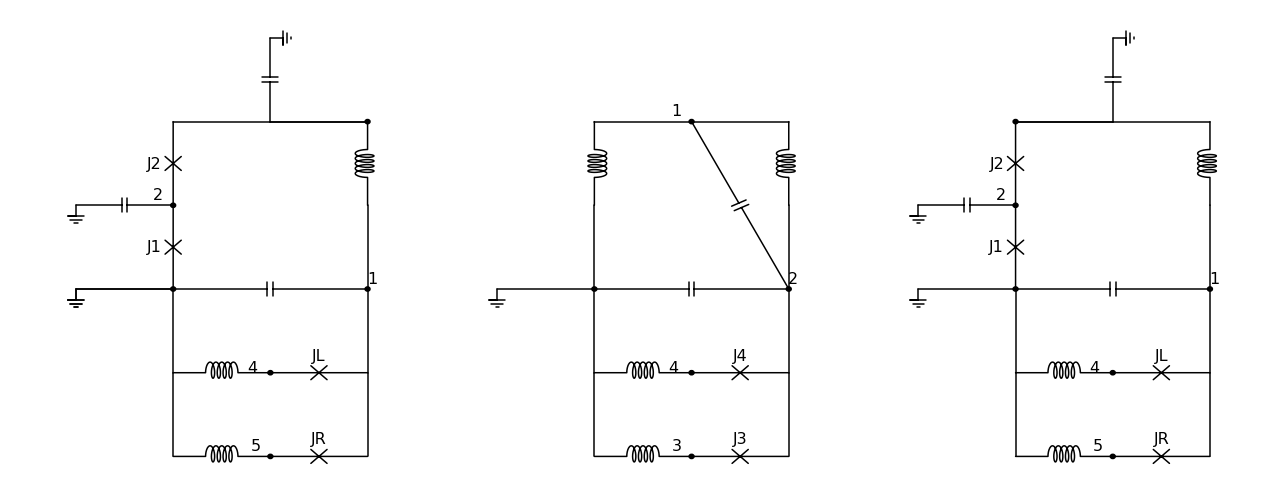}
         \label{fig:Josephson_circuit_model}
     \end{subfigure}
     \hfill
     \begin{subfigure}[b]{0.8\columnwidth}
         \centering
         \caption{}
         \includegraphics[width=1\columnwidth]{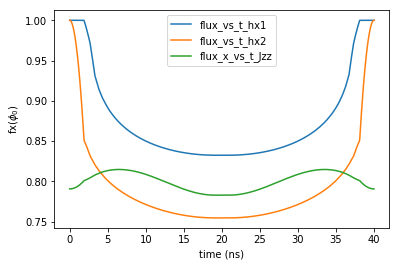}%
         \label{fig:fx_schedule}
     \end{subfigure}
     \hfill
     \begin{subfigure}[b]{0.8\columnwidth}
         \centering
         \caption{}
         \includegraphics[width=1\columnwidth]{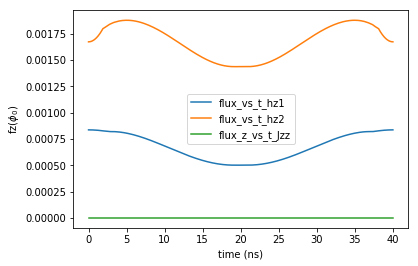}%
         \label{fig:fz_schedule}
     \end{subfigure}
     \hfill
     \begin{subfigure}[b]{0.8\columnwidth}
         \centering
         \caption{}
         \includegraphics[width=1\columnwidth]{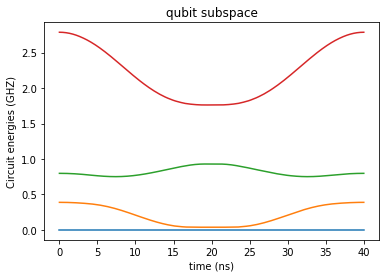}%
         \label{fig:levels_Josephson_circuit}
     \end{subfigure}     
     
        \caption{Flux schedule and levels evolution of Josephson circuits for realizing the two-spin gate (a)Structure of the Josephson circuits used for realizing our two-spin gadget (b)Flux bias schedule (x loop flux bias vs time) for the two-spin gate (c)Flux bias schedule (z loop flux bias vs time) for the two-spin gate (d)The evolution of energy levels of the Josephson circuits caused by the flux bias schedule for the two-spin gate }
        \label{fig:flux schedule of two-qubit gate}
\end{figure}

For our dimer DQA gadget, the appropriate circuit model flux bias schedules that give rise to the desired spin model schedules have already been published in our previous paper\cite{translator1}. For our two-qubit gate, the flux bias schedules corresponding to the optimized two-qubit CNOT-like gate are shown in Fig.\ref{fig:flux schedule of two-qubit gate}. Fig.\ref{fig:flux schedule of two-qubit gate}(a) shows the structure of the simple Josephson circuits used in the calculation. Fig.\ref{fig:flux schedule of two-qubit gate}(b) and (c) show the translated flux bias schedules of the Josephson circuits. The evolution of the lowest 4 circuit energy levels during the gate operation is shown in Fig.\ref{fig:flux schedule of two-qubit gate}(d), which reproduces the desired evolution of levels in the spin model (Fig.\ref{fig:two-qubit gate}(c)). The translated flux bias control indicates that our two-qubit gate can be realized by using only flux bias control. 

In the Josephson circuits settings, the dominating coupling between qubits is from the inductive coupling between the persistent currents. The advantage of this coupling scheme is that it will not compromise the coherence of the flux qubit when the coupling strength is increased. For the new variety of flux qubit-- fluxonium, the coherence is enhanced. The high coherence is rooted in the suppressed transition matrix element of the charge operator of the fluxonium\cite{bao2021fluxonium}. However, for gates based on capacitive coupling (which is the dominating design for gate-based quantum computation), this suppression will inevitably slows down the gate speed. The inductive coupling, however, doesn't suffer from the above limitation\cite{bao2021fluxonium}. Therefore, inductive coupling resolves the limitation caused by the capacitive coupling and potentially can lead to much higher gate speed. The inclusion of tunable coupler in our Josephson circuits is useful for realizing large scale quantum operation in flux qubit systems, since the coupling between active part and rest of the system can be turned off during gate operation. A tunable coupler is crucial for maintaining high-fidelity for parallel operations\cite{bao2021fluxonium}. The circuit translation here can also be readily applied to quantum computer system with CSFQ being replaced by other variants of flux qubits such as fluxonium, whose circuit Hamiltonian is basically the same as the example here.

\section{Summary and Discussion}
The faster speed and operational convenience of two-qubit gate with flux bias control makes it an important candidate for future large-scale quantum computers based on high coherence flux qubits. Based on a properly designed two-spin gadget which has small gaps during the evolution of energy levels, we build a CNOT-equivalent gate which can reach a fidelity larger than $99.9\%$ within 40ns. Moreover, we also use the Schrieffer-Wolff Transformation to translate the spin model Ising coefficients schedule to circuit model flux bias schedule for realistic flux qubit circuits coupled by a tunable rf-squid. The two-qubit entangling gate scheme introduced here is suitable for the large scale flux qubit systems dominated by inductive couplings. Comparing with the current gate-based quantum computation systems dominated by capacitive couplings, it can resolve the conflict between the speed and a high coherence.

Additionally, besides the usefulness in the gate-based quantum computation, our gate could also be used for preparing an excited initial state for the diabatic reverse annealing (DRA), which is the most promising annealing scheme for showing an exponential quantum speedup\cite{Crosson2020}. To realize a diabatic annealing, one way is to start from the ground state and introduce diabatic transitions across small gaps via a short evolution time,  the other way is to initialize the system in an excited state and let the system de-excite to the gnd state through small gaps during the evolution\cite{Crosson2020}. Our gate can be used to prepare an excited state for the diabatic reverse annealing using only flux bias control, which is the only control available in most of annealers.

\begin{acknowledgments}

The research is based upon work supported by IQC, University of Waterloo.

\end{acknowledgments}

\bibliography{two_spin_gadget_gate}

\end{document}